\documentclass[twocolumn,showpacs,preprintnumbers,amsmath,amssymb]{revtex4}
 \usepackage{graphicx}  

\newcommand{\beq}{\begin{eqnarray}}
\newcommand{\eeq}{\end{eqnarray}}

\begin{document}

\title{Neutron-star radii based on realistic nuclear interactions}

\author{Y.\ Yamamoto$^{1}$}
\email{ys_yamamoto@riken.jp}
\author{H.\ Togashi$^{1}$$^{2}$}
\author{T.\ Tamagawa$^{1}$}
\author{T.\ Furumoto$^{3}$}
\author{N.\ Yasutake$^{4}$}
\author{Th.A.\ Rijken$^{5}$$^{1}$}
\affiliation{
$^{1}$RIKEN Nishina Center, 2-1 Hirosawa, Wako, 
Saitama 351-0198, Japan\\
$^{2}$Research Institute for Science and Engineering, Waseda University, Okubo, 
Shinjuku-ku, Tokyo, 169-8555, Japan\\
$^{3}$Graduate School of Education, Yokohama National
University, Yokohama 240-8501, Japan\\
$^{4}$Department of Physics, Chiba Institute of Technology, 2-1-1 Shibazono
Narashino, Chiba 275-0023, Japan\\
$^{5}$IMAPP, University of Nijmegen, Nijmegen, The Netherlands
}

%

\begin{abstract}
The existence of neutron stars with $2M_\odot$ requires the
strong stiffness of the equation of state (EoS) of neutron-star matter.
We introduce a multi-pomeron exchange potential (MPP) working universally 
among 3- and 4-baryons to stiffen the EoS. Its strength is restricted 
by analyzing the nucleus-nucleus scattering with the G-matrix folding model. 
The EoSs are derived using the Brueckner-Hartree-Fock (BHF) and the 
cluster variational method (CVM) with the nuclear interactions ESC and AV18.
The mass-radius relations are derived by solving the 
Tolmann-Oppenheimer-Volkoff (TOV) equation, where the maximum masses over 
$2M_\odot$ are obtained on the basis of the terrestrial data.
Neutron-star radii $R$ at a typical mass $1.5M_\odot$ are predicted
to be $12.3\!\sim\!13.0$ km. The uncertainty of calculated radii is mainly 
from the ratio of 3- and 4-pomeron coupling constants, which cannot be fixed
by any terrestrial experiment.
Though values of $R(1.5M_\odot)$ are not influenced by hyperon-mixing effects,
finely-observed values for them indicate degrees of EoS softening by
hyperon mixing in the region of $M\!\sim\!2M_\odot$.
If $R(1.5M_\odot)$ is less than about 12.4 km,
the softening of EoS by hyperon mixing has to be weak.  
Useful information can be expected by the space mission NICER offering 
precise measurements for neutron-star radii within $\pm 5\%$.
\end{abstract}


\maketitle

\parindent 15 pt

\section{Introduction}

In studies of neutron stars, the fundamental role is played by
the equation of state (EoS) for dense nuclear matter.
The observed masses of neutron stars J1614$-$2230~\cite{Demorest10} and
J0348+0432~\cite{Antoniadis13} are given as $(1.97\pm0.04)M_{\odot}$ 
and $(2.01\pm0.04)M_{\odot}$, respectively, being severe conditions 
for the stiffness of EoS of neutron-star matter.
It is well known that the stiff EoS giving the maximum mass of 
$2M_{\odot}$ can be derived from the existence of strongly 
repulsive effects in the high-density region. 
In the non-relativistic approaches, three-body repulsions (TBR) interactions 
among nucleons are taken into account. 
In \cite{Gandolfi12}, for instance, neutron matter EoSs and 
mass($M$)$-$radius($R$) relations of neutron stars were studied 
using quantum Monte Carlo technique with the AV8' interaction 
added by $3n$ repulsions.
In relativistic mean field (RMF) models, self-interactions of repulsive 
vector mesons are taken into account to stiffen EoSs. 

On the other hand, hyperon ($Y$) mixing in neutron-star matter 
brings about remarkable softening of the EoS, canceling the TBR effect 
for the maximum mass~\cite{Baldo00,Vidana00,NYT}:
With increasing of baryon density toward centers of neutron stars,
chemical potentials of neutrons become high so that neutrons at
Fermi surfaces are changed to hyperons ($Y$) via strangeness non-conserving
weak interactions overcoming rest masses of hyperons.
This hyperon mixing to neutron-star matter exists by all means.
One of the ideas to avoid this ``Hyperon puzzle in neutron stars" is to 
assume that the many-body repulsions work universally for every kind of baryons.
In Refs.~\cite{YFYR13,YFYR14,YFYR15}, the multi-pomeron exchange potential
(MPP) was introduced as a model of universal repulsions among three and 
four baryons on the basis of the Extended Soft Core (ESC) $B\!B$
interaction model developed by two of authors (T.R. and Y.Y.) 
and M.M. Nagels~\cite{ESC08,ESC08c1}.
MPP and the additional three-body attraction (TBA) were restricted on the 
basis of terrestrial experiments, where another adjustable parameter
was not used for the stiffness of EoS.

Comparing to the measurement of neutron-star masses,
the observational determination of their radii has been difficult.
Though the radius can be extracted from the analysis of X-ray spectra
emitted by the neutron star atmosphere, very different values for 
stellar radii have been derived because of uncertainties of
the modeling of the X-ray emission. Now, by the space mission 
NICER (Neutron star Interior Composition ExploreR)~\cite{NICER},
high-precision X-ray astronomy is expected to offer precise measurements
for neutron-star radii within $\pm 5\%$.

In this work, we start from the EoS of neutron-star matter with 
baryonic constituents, not considering a possible transition to 
deconfined quark matter, and derive basic $M\!R$ relations 
by solving the Tolmann-Oppenheimer-Volkoff (TOV) equation.
Our EoS is derived from the realistic baryon-baryon ($B\!B$) interaction model
added by MPP and TBA,  This interaction model is confirmed by
rich terrestrial data, and then $M\!R$ relations can be 
predicted within a small range. 
As shown later, the important point in this work is that basic features of 
$M\!R$ relations are determined substantially by MPP parts with minor
contributions from TBA parts.

The modern $N\!N$ interaction models, one of which is our ESC, are constructed
with high accuracy in reproducing $N\!N$ scattering data. It is well known,
however, that these potentials lead to different saturation curves
($E/A$ values as a function of density), and these curves are controlled 
mainly by tensor components included in interaction models. 
It is interesting to study how the difference of interaction models
has an effect on the $M\!R$ relations.
In this work, we pick up the AV18 potential~\cite{AV18}
as an example giving the far shallower saturation curve than ESC.

In our approach, no ad hoc parameter is included to control the stiffness 
of neutron-star EoS. This means that we can predict radii of neutron stars 
as a function of their masses, which should be confirmed 
by the coming observational data.
We adopt the Burueckner-Hartree-Fock (BHF) theory in order to treat 
baryonic many-body systems with realistic $B\!B$ interaction models,
and study properties of baryonic matter including not only nucleons 
but also hyperons with use of the lowest-order G-matrix theory with 
the continuous choice for intermediate single particle potentials. 
Methods of G-matrix calculations in this work
are the same as those in \cite{YFYR13,YFYR14,YFYR15}, but
numerical results of G-matrices are different from those
in these previous works because the computation program is improved.

The EOSs and $M\!R$ relations by BHF are compared with those calculated 
with the cluster variational method (CVM)~\cite{Togashi1,Togashi2,Togashi3}
to discuss the theoretical uncertainties in predicted values of
neutron-star radii with respect to the many-body approaches.


This paper is organized as follows:
In Sect.II, on the basis of realistic interaction models,
the EoSs and the $M\!R$ relations of neutron stars are derived:
In IIA, the MPP and TBA parts are restricted so as to reproduce
the angular distributions of $^{16}$O+$^{16}$O scattering and
nuclear saturation properties.
In IIB, the EoSs of $\beta$-stable neutron-star matter are
derived with use of BHF. The $M\!R$ relations are obtained
by solving the TOV equation.
In IIC, the EoSs and $M\!R$ relations are investigated
with use of CVM.
In Sec.III, the EoSs of hyperonic nuclear matter are derived
and effects of hyperon mixing on the $M\!R$ relations are
investigated.
In Sec.IV, our predictions for values of $R(1.5M_\odot)$
are summarized.

\section{Baryon-Baryon interaction and neutron-star EoS}

\subsection{Many-body repulsion}

We start from the ESC $B\!B$ interaction.
The latest version of ESC is named as ESC08c~\cite{ESC08c1}. 
Hereafter, ESC means this version.
In this modeling, important parts of $B\!B$ repulsive cores are described
by pomeron exchanges. They can be extended straightforwardly to
$N$-body repulsions by multi-pomeron exchanges, called here
the multipomeron-exchange potential(MPP) \cite{YFYR13,YFYR14}.
The $N$-body local potential $ W^{(N)}$ by pomeron exchange is
\begin{eqnarray}
&& W^{(N)}({\bf x}_1, ..., {\bf x}_N) = g_P^{(N)} g_P^N\ \left\{
\int\frac{d^3k_i}{(2\pi)^3} e^{-i{\bf k}_i\cdot{\bf x}_i}\right\}
\nonumber\\ 
&& \times (2\pi)^3\delta(\sum_{i=1}^N {\bf k}_i)
\Pi_{i=1}^N \left[\exp\left(-{\bf k}_i^2\right)\right]\cdot {\cal M}^{4-3N} ,
\label{eq:1}
\end{eqnarray}
where $g_P$ and $g_P^{(N)}$ are two-body and $N$-body pomeron strengths, 
respectively, and the (low-energy) pomeron propagator is the same as 
the one used in the two-body pomeron potential.
The defined MPP works universally among baryons
because the pomeron is an SU(3)-singlet in flavor (and color) space. 
The effective two-body potential in a baryonic medium is obtained
by integrating over the coordinates ${\bf x}_3,..., {\bf x}_N$ 
as follows:
\begin{eqnarray}
&& V_{eff}^{(N)}({\bf x}_1,{\bf x}_2) 
 \nonumber \\
&& 
 = \rho_{}^{N-2} 
 \int\!\! d^3\!x_3 ... \int\!\! d^3\!x_N\ 
 W^{(N)}({\bf x}_1,{\bf x}_2, ..., {\bf x}_N)
 \nonumber \\
&& 
=g_P^{(N)} g_P^N\frac{\rho^{N-2}}{{\cal M}^{3N-4}}
 \left(\frac{m_P}{\sqrt{2\pi}}\right)^3
 \exp\left(-\frac{1}{2}m_P^2 r_{12}^2\right).
\label{eq:2}
\end{eqnarray}
Now, we assume that
the dominant mechanism is triple and quartic pomeron exchange.
The values of the two-pomeron strength $g_P$ and 
the pomeron mass $m_P$ are the same as those in ESC.
The scale mass ${\cal M}$ is taken as the proton mass.

Because MPP is purely repulsive, it is considered generally to 
add a three-body attraction (TBA) to ESC together with MPP 
in order to reproduce the nuclear saturation property.
We introduce here a phenomenological potential represented 
as a density-dependent two-body interaction
\begin{eqnarray}
V_{A}(r;\rho)= V_0\, \exp(-(r/2.0)^2)\, \rho\, 
\exp(-\eta \rho)\, (1+P_r)/2 \ .
\label{eq:4}
\end{eqnarray}
$P_r$ is a space-exchange operator so that
$V_{A}(r;\rho)$ works only in even states due to $(1+P_r)$.
$V_0$ and $\eta$ are treated as adjustable parameters.

It is evident here that MPP is defined as a straight forward extension
of the ESC modeling. However, because its strength is restricted
on the basis of experimental data, it is meaningful that
as a phenomenological model the same form of MPP is added to AV18 
together with the $V_{A}$.

As shown in ref.\cite{FSY}, the repulsive effects by MPP in nucleon sectors 
appear in angular distributions of $^{16}$O+$^{16}$O elastic scattering 
at an incident energy per nucleon $E_{in}/A=70$ MeV, $etc$.
Such a scattering phenomenon can be analyzed quite successfully with the complex 
G-matrix folding potentials derived from free-space $N\!N$ interactions:
G-matrix calculations are performed in nuclear matter, 
and G-matrix interactions are represented in coordinate space to construct 
nucleus-nucleus folding potentials~\cite{FSY}. 
In the same way as \cite{YFYR13,YFYR14,YFYR15}, the analyses for the 
$^{16}$O$+^{16}$O elastic scattering at $E_{in}/A=70$ MeV are performed.
The MPP strengths ($g_P^{(3)}$ and $g_P^{(4)}$), together with $V_{A}$,
are adjusted to reproduce the scattering data using the G-matrix folding potential
together with the condition that the saturation parameters 
of nuclear matter are reproduced well.
Backward angular distributions of $^{16}$O$+^{16}$O scattering are substantially 
restricted by the MPP repulsive contributions in the density region
of $\rho_0<\rho<2\rho_0$~\cite{FSY14}.
They are not so dependent on a ratio of contributions of triple 
and quartic pomeron exchanges, and we can find various combinations
of $g_P^{(3)}$ and $g_P^{(4)}$ reproducing the data equally well.
As found in Eq.(\ref{eq:2}), the contributions from triple and
quartic components are proportional to $\rho$ and $\rho^2$, respectively. 
Therefore, the latter contribution plays a remarkable role to stiffen 
the EoS in high density region. 

\begin{table}[tbh]
\centering 
\caption{Parameters $g_P^{(3)}$ and $g_P^{(4)}$ included in MPP}
\label{tab:1}       
\vskip 0.2cm
\begin{tabular}{lcc}
\hline\noalign{\smallskip}
&\ $g_P^{(3)}\ $ &\ $g_P^{(4)}$ \ \\ 
\noalign{\smallskip}\hline\noalign{\smallskip}
MPa     & 2.62 & 40.0  \\
MPb     & 3.37 & 0.0   \\
MPa$^+$ & 1.84 & 80.0  \\
\noalign{\smallskip}\hline
\end{tabular}
\end{table}

The chosen parameter sets are listed in Table \ref{tab:1}, 
where the parameter values for these sets are different from 
those in \cite{YFYR14,YFYR15} because of the improvement of
G-matrix calculations. For TBA parts in the case of using ESC, 
we take $V_0=-8.0$ MeV and $\eta =4.0$ fm$^{-1}$ in three sets 
(MPa, MPb, MPa$^+$) of $g_P^{(3)}$ and $g_P^{(4)}$.
Although another choice of $\eta$ leads mainly to a change in the
saturation densities, its impacts on $M\!R$ curves of neutron stars
are small. For simplicity, the value of $\eta=4.0$ fm$^{-1}$
is fixed in this paper.
In the case of using AV18, we take $V_0=-35.0$ MeV 
($\eta =4.0$ fm$^{-1}$) for the TBA part with the same strengths 
of MPP, where the deeper value of $V_0$ in the AV18 case is needed 
to reproduce the saturation properties.
MPa and MPb are denoted as MPa' and MPb', respectively,
when $V_0=-8.0$ MeV in the formers are changed to $V_0=-35.0$ MeV.
Thus, ESC (AV18) combined with MPa (MPa') 
is denoted as ESC+MPa (AV18+MPa'), and so on.

\begin{figure}[htb!]
\includegraphics[width=7cm]{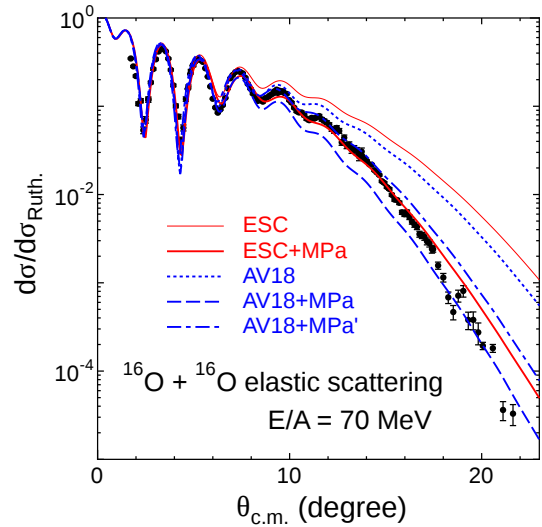}
\vskip 3cm
\caption{(Color online)
Differential cross sections for $^{16}$O+$^{16}$O elastic 
scattering at $E_{in}/A=70$ MeV calculated with G-matrix folding potentials.
Solid, dashed and dot-dashed curves are for ESC+MPa, AV18+MPa and
AV18+MPa', respectively. Dotted curves are for ESC and AV18.
}
\label{xsO16O16}
\end{figure}

\begin{figure}[htb]
\includegraphics[width=\columnwidth]{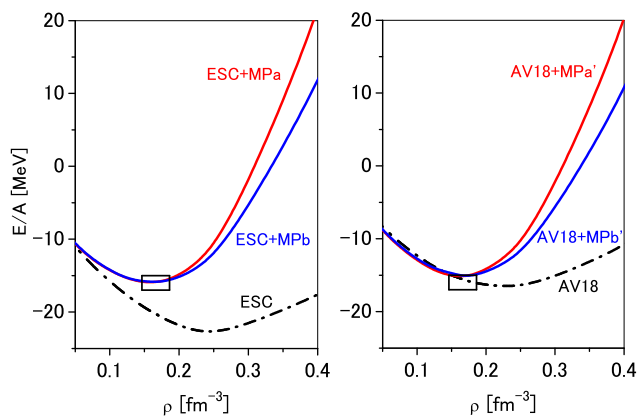}
\caption{(Color online) Energy per particle $E/A$ as a function of 
nucleon density $\rho$ symmetric matter. 
Solid curves in the left (right) panel are obtained by 
ESC+MPa and ESC+MPb (AV18+MPa' and AV18+MPb'), and 
the dot-dashed one is by ESC (AV18).
The box shows the empirical value. 
}
\label{saturation}
\end{figure}

Let us show differential cross sections for 
$^{16}$O+$^{16}$O elastic scattering at $E_{in}/A=70$ MeV calculated 
with the G-matrix folding potentials in comparison with 
the experimental data~\cite{Nuoffer}. 
In comparing the G-matrix folding potentials derived from
ESC and AV18, it should be noted that not only their real parts
but also their imaginary parts are different from each other.
As shown by the $E/A$ curves in Fig.~\ref{saturation},
the real part for AV18 is shallower than that for ESC.
Furthermore, the imaginary potential for AV18 is considerably weaker
than for ESC. Both features can be understood by the fact that
the weight of the tensor component in AV18 is larger than 
in the case of ESC:
Tensor-force contributions are suppressed more efficiently in medium
than central-force ones, which leads to less attractive G-matrices.
In the cases of using ESC, the derived imaginary potentials
are so suitable to reproduce the $^{16}$O+$^{16}$O scattering data
with no adjustment. On the other hand, the factor $N_W=1.5$ is
multiplied on the imaginary potentials derived from AV18.
In Fig.~\ref{xsO16O16}, thin solid and dotted curves are obtained
with ESC and AV18, respectively, which deviate substantially from the data.
The solid curve is for ESC+MPa, fitting the data nicely. The dashed curve 
is for AV18+MPa with $V_0=-8.0$ MeV, which demonstrates that the 
$^{16}$O$+^{16}$O folding potential is too repulsive to reproduce the data.
The dot-dashed curve is for AV18+MPa', in which we take $V_0=-35.0$ MeV
without changing the MPP strength.
Thus, in the AV18 case, it is necessary to make $V_{A}$ more
attractive than that in the ESC case in order to reproduce 
the data well.  

Similar curves can be obtained in the case of using ESC+MPb,
AV18+MPb and AB18+MPb', where MPb and MPb' include the TBA parts 
of $V_0=-8.0, -35.0$ MeV, respectively.

In Fig.~\ref{saturation}, 
we show the energy curves of symmetric nuclear matter, namely 
binding energy per nucleon  $E/A$ as a function of density. 
Solid curves in the left (right) panel are obtained by 
ESC+MPa and ESC+MPb (AV18+MPa' and AV18+MPb'), and 
the dot-dashed one is by ESC (AV18).
The box in the figure shows the area where nuclear 
saturation is expected to occur empirically.
The minimum value of $E/A$ for AV18 is found to be
$-16.5$ MeV at $\rho=0.229$ fm$^{-3}$, being considerably
shallower than that for ESC $-22.5$ MeV at $\rho=0.255$ fm$^{-3}$.
This difference of $E/A$ values for ESC and AV18 is related to
the necessity of taking different values for $V_0$
($-8.0$ and $-35.0$ MeV for ESC and AV18, respectively).

\begin{table}[tbh]
\centering 
\caption{Calculated values of saturation parameters.}
\label{tab:2}       
\vskip 0.2cm
\begin{tabular}{lccccc}
\hline\noalign{\smallskip}
& $\rho_0$ & $E/A$ & $E_{sym}$ & $L$ & $K$ \\ 
& (fm$^{-3}$) & (MeV) & (MeV) & (MeV) & (MeV) \\
\noalign{\smallskip}\hline\noalign{\smallskip}
ESC+MPa     & 0.151 & $-$16.3  & 31.7 & 55.7 & 248  \\
ESC+MPb     & 0.155 & $-$16.1  & 31.4 & 49.2 & 217  \\
ESC+MPa$^+$ & 0.148 & $-$16.5  & 31.4 & 55.1 & 275  \\
AV18+MPa'   & 0.159 & $-$15.3  & 29.4 & 50.5 & 263  \\
AV18+MPb'   & 0.165 & $-$15.2  & 29.9 & 52.0 & 234  \\
\noalign{\smallskip}\hline
\end{tabular}
\end{table}

The EoS is specified by the following quantities:
The difference between the $E/A$ curves for neutron matter and
symmetric matter gives the symmetry energy $E_{sym}(\rho)$, and
its slope parameter is defined by 
$L=3\rho_0 \left[\frac{\partial E_{sym}(\rho)}{\partial \rho}\right]_{\rho_0}$.
The incompressibility is defined by
$K=9\rho_0^2 \left[\frac{\partial^2}{\partial \rho^2} E/A(\rho)\right]_{\rho_0}$.
Calculated values of these quantities at saturation density $\rho_0$ are summarized 
in Table \ref{tab:2}. The minimum values of $E/A$ curves in all cases 
turn out to be close to the empirical value. 
The symmetric energies and their slope parameters
are similar to each other and consistent with the empirical indications.
The difference among these sets appears in the values of the incompressibility $K$.
The experimental values of $K$ are given as 
$220 \sim 250$ MeV~\cite{Sagawa}. 
The value of $K=275$ MeV at 
$\rho_0=0.148$ fm$^{-3}$ in the MPa$^+$ case seems to be  too large 
in comparison with the experimental indication.
In the following part of this paper, MPa$^+$ is used only to demonstrate
the relation between the MPP repulsion and the stiffness of EoS.

\subsection{EoS and $M\!R$ relation}

Using our interaction models,
we derive the EoS of $\beta$-stable neutron-star matter composed of
neutrons ($n$), protons ($p$), electrons ($e^-$), muons ($\mu^-$),
The EoSs obtained from G-matrix calculations
are used for $\rho > 0.24$ fm$^{-3}$.
Below 0.12 fm$^{-3}$ we use the EoS of the crust obtained
in \cite{Baym1,Baym2}. Then, the EoSs for $\rho > 0.24$ fm$^{-3}$
and $\rho < 0.12$ fm$^{-3}$ are connected smoothly.
The $E/A$ curves obtained from G-matrix calculations are fitted
by analytical functions, giving rise to analytical expressions
for energy density, chemical potential and pressure.

\begin{figure}[htb]
\includegraphics[width=\columnwidth]{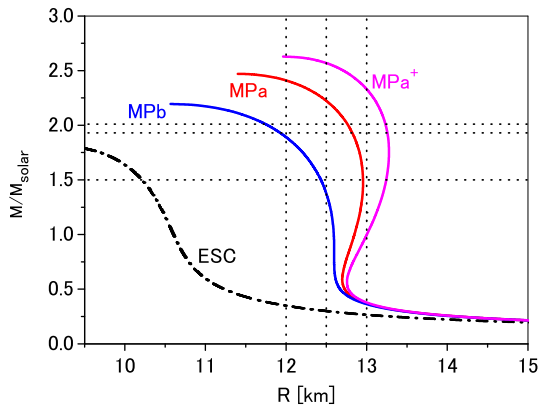}
\caption{(Color online)
Neutron-star masses as a function of the radius $R$.
Solid, dashed and dotted curves are for MPa, MPa$^+$ and MPb,
respectively. Dot-dashed curve is for ESC.
Upper two dotted horizontal lines show the observed masses
$1.97M_\odot$ and $2.01M_\odot$ of J1614-2230 and J0348+0432,
respectively. Lower dotted line shows mass $1.5M_\odot$ of
a typical neutron star.}
\label{MRnuc1}
\end{figure}

Assuming a mixed matter of $n$, $p$, $e^-$ and $\mu^-$
in chemical equilibrium, we solve the TOV equation for the hydrostatic 
structure to obtain mass-radius relations of neutron stars.
In Fig.\ref{MRnuc1}, let us demonstrate the obtained $M\!R$ relations
of neutron stars.  Solid curves are for ESC+MPa/MPb/MPa$^+$,and the 
dot-dashed one for ESC. The EoSs including MPP repulsions are found 
to be stiff enough to give 2$M_{\odot}$ maximum masses. 
It should be noted that the 2$M_{\odot}$ masses are obtained
with no ad hoc parameter to stiffen EoSs in our approach, 
because our MPP repulsions are restricted so as to reproduce the 
$^{16}$O+$^{16}$O scattering data. It can be checked, here, that 
contributions of TBA to the $M\!R$ curves are small in the case
of using ESC, demonstrating that basic features of $M\!R$ relations 
are mainly determined by MPP contributions.
The difference between MPa (MPa$^+$) and MPb is due to the 
quartic-pomeron exchange term included in the formers. 
The strength of the effective two-body interaction derived 
from the quartic-pomeron exchange is proportional to $\rho^2$, 
and the contribution becomes sizeable in the 
high-density region, making the maximum mass so large.
Here, the important point in Fig.\ref{MRnuc1} is as follows:
The repulsive effect by MPP is shown symbolically as 
MPa$^+$ $>$ MPa $>$ MPb, and the increase of the repulsive effect
raises both $M$ and $R$ of a neutron star. 
In our present approach, only the strength of this repulsive effect 
plays a role to adjust the stiffness of the EoS. 
This effect changes both mass and radius of a neutron star simultaneously:
There is no parameter changing mass and radius independently.
Now, let us remark the obtained values of $R(1.5M_\odot)$,
being 12.3 km, 12.9 km and 13.1 km for MPb, MPa and MPa$^+$, respectively.

\begin{figure}[htb]
\includegraphics[width=\columnwidth]{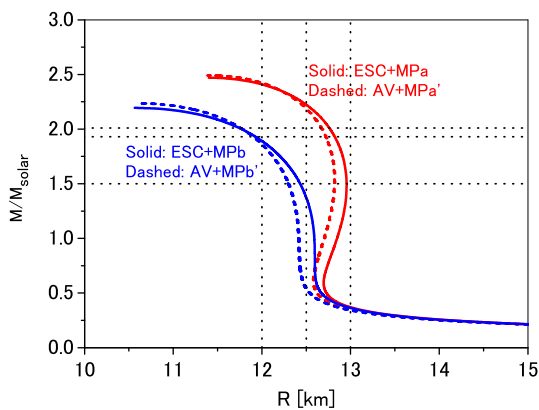}
\caption{(Color online)
Neutron-star masses as a function of the radius $R$.
Solid curves are for ESC+MPa and ESC+MPb.
Dashed curves are for AV18+MPa' and AV18+MPb'.
Also see the caption of Fig.\ref{MRnuc1}.
}
\label{MRnuc2}
\end{figure}

In Fig.\ref{MRnuc2}, the $M\!R$ curves for AV18 are compared with
those for ESC. Solid curves are for ESC+MPa and ESC+MPb,
being the same as the corresponding curves in Fig.\ref{MRnuc1}.
The dashed curves are for AV18+MPa' and AV18+MPb', 
n which $V_0=-8.0$ MeV in MPa/MPb is changed to the more attractive
value of $V_0=-35.0$ MeV so as to reproduce the $^{16}$O+$^{16}$O 
scattering data.
The value of $R(1.5M_\odot)$ for AV18+MPa' (AV18+MPb') is found 
to be smaller by 0.2 km (0.1 km) than that for ESC+MPa (ESC+MPb).

\subsection{Calculations by the variational method}

The variational method is another powerful method for neutron-star EoS 
in the non-relativistic approach. In Refs.\cite{Togashi1,Togashi2,Togashi3},
one of the present authors (H.T.) and his collaborators developed 
a cluster variational method (CVM) for uniform nuclear matter 
with arbitrary proton fractions.
It is important to compare the BHF results in the previous sections with 
those by CVM using the same interaction models. 
We adopt here AV18+MPa' and AV18+MPb'.
%
In Table \ref{tab:3}, we show the values
of saturation parameters calculated by CVM.
In comparison with the BHF results for the same interactions
in Table \ref{tab:2}, CVM turns out to give shallower values 
of $E/A$ by about 2 MeV than BHF.

\begin{table}[tbh]
\centering 
\caption{Values of saturation parameters calculated by CVM.
}
\label{tab:3}       
\vskip 0.2cm
\begin{tabular}{lccccc}
\hline\noalign{\smallskip}
       & $\rho_0$ & $E/A$ & $E_{sym}$ &  $L$ & $K$  \\ 
& (fm$^{-3}$) & (MeV) & (MeV)  & (MeV) & (MeV)  \\
\noalign{\smallskip}\hline\noalign{\smallskip}
AV18+MPa'  & 0.155 & $-$12.85 & 26.6 & 47 & 275  \\
AV18+MPb'  & 0.164 & $-$12.92 & 27.5 & 48 & 252 \\
\noalign{\smallskip}\hline
\end{tabular}
\end{table}

\begin{figure}[htb!]
\includegraphics[width=8cm]{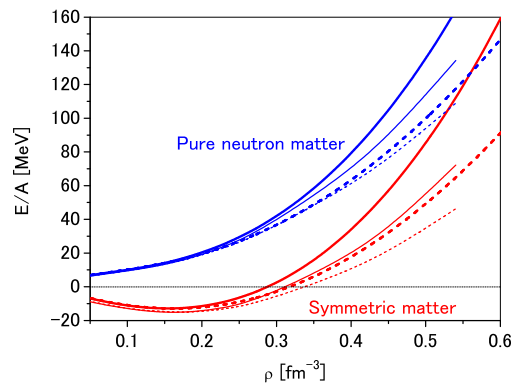}
\caption{(Color online)
$E/A$ curves of symmetric and pure neutron matter.
Solid and dashed curves are for AV18+MPa' and AV18+MPb', respectively.
Thick (thin) curves are obtained by CVM (BHF).}
\label{eneAV}
\end{figure}

Fig.\ref{eneAV} shows the $E/A$ curves of symmetric and pure neutron matter
obtained by CVM.
The solid and dashed curves are for AV18+MPa' and AV18+MPb', respectively.
The corresponding $E/A$ curves obtained by BHF are drawn by thin solid
and dashed curves for AV18+MPa' and AV18+MPb', respectively.
Here, the CVM results are found to become more repulsive than 
the BHF ones with increasing of density, especially in symmetric matter.
This means that strong repulsive interactions are evaluated larger
by CVM than BHF. It is difficult to determine which is reasonable.
In \cite{Baldo12}, the BHF results for the EoS of nuclear matter
were compared with the results by other many-body methods
using Argonne-type interactions. They found that the formers were
significantly different from the latters in symmetric matter.
Such situations are similar to our present case. Thus, it is considered 
that there still remains an important problem how to obtain the most 
realistic description of nuclear-matter EoS.

\begin{figure}[htb!]
\includegraphics[width=8cm]{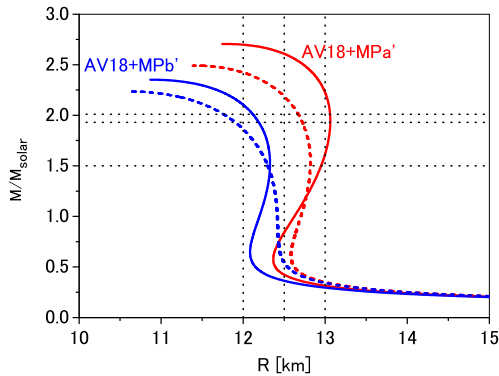}
\caption{(Color online)
$M\!R$ curves of neutron stars for AV18+MPa' and AV18+MPb'.
Solid and dashed curves are obtained by CVM and BHF, respectively.
Also see the caption of Fig.\ref{MRnuc1}.}
\label{NScvm}
\end{figure}

In Fig.\ref{NScvm}, the solid curves show the $M\!R$ relations
obtained by CVM for AV18+MPa' and AV18+MPb', where
the EoS in the crust region is treated in the same way as
the previous section, being different from the treatment in \cite{Togashi2}.
For comparison, here, the BHF results for AV18+MPa' and AV18+MPb'
are shown by dashed curves.
Though the masses and radii obtained by CVM are larger than those by BHF
in the case of using the same interaction, the difference of 
$M(1.5M_\odot)$ values is less than 0.2 km.

The reason of different $M\!R$ curves is because the MPP repulsions
are evaluated larger by CVM than BHF. Considering that the MPP strength
is determined in the BHF treatment (G-matrix folding model),
it is reasonable to reduce the MPP strength in the CVM treatment
so as to reproduce $E/A$ values properly. Then, it is expected that
the $M\!R$ curve for CVM becomes close to that for BHF.

\begin{table}[tbh]
\centering 
\caption{Calculated values of $R(1.5M_\odot)$.}
\label{tab:4}       
\vskip 0.2cm
\begin{tabular}{l|c|c}
\hline\noalign{\smallskip}
model &method & $R(1.5M_\odot)$ [km] \\ 
\noalign{\smallskip}\hline\noalign{\smallskip}
ESC+MPa   & BHF   &  13.0 \\
AV18+MPa' & BHF   &  12.8 \\
          & CVM   &  13.0 \\
\noalign{\smallskip}\hline\noalign{\smallskip}
ESC+MPa$^+$ & BHF &  13.2 \\
\noalign{\smallskip}\hline\noalign{\smallskip}
ESC+MPb   & BHF   &  12.4 \\
AV18+MPb' & BHF   &  12.3 \\
          & CVM   &  12.3 \\
\noalign{\smallskip}\hline
\end{tabular}
\end{table}

In Table \ref{tab:4}, we summarize the calculated values
of $R(1.5M_\odot)$.
Here, it is notable that the value of $R(1.5M_\odot)$=12.3 km 
for AV18+MPb' is quite similar to that for ESC+MPb, and
this value for AV18+MPb' is reproduced by both of BHF and CVM.
On the other hand, in the corresponding three cases including MPa,
the values of $R(1.5M_\odot)$ are slightly different from each other.
As the MPP strength becomes more repulsive, the value of $R(1.5M_\odot)$
obtained by CVM becomes larger than that by BHF.

Thus, considering the ambiguities of interactions and methods, 
we can say as follows:
For the MPb-type interactions including 3-body repulsion only,
we expect $R(1.5M_\odot)= 12.3\!\sim\!12.4$ km. 
For the MPa-type interactions including 3- and 4-body repulsions, 
we expect $R(1.5M_\odot)= 12.8\!\sim\!13.0$ km.
As shown in next section, these results for $R(1.5M_\odot)$
are not changed by effects of hyperon mixing.

\section{Hyperon mixing}

Let us recapitulate our method to derive the EoS of baryonic matter 
composed of nucleons ($N=n,p$) and hyperons ($Y=\Lambda, \Sigma^-$).
A single particle potential of $B$ particle in $B'$ matter
$U_{B}^{(B')}(k)$ is given by summing up G-matrix elements 
$\langle kk'|G_{BB',BB'}|kk'\rangle$ 

\begin{eqnarray}
U_{B}^{(B')}(k)&=& \sum_{k',,k_F^{(B')}} \langle kk'|G_{BB',BB'}|kk'\rangle
\end{eqnarray}
with $B,B'=N,Y$, where spin isospin quantum numbers are implicit.
Then, a single particle potential of $B$ in baryonic matter is given by
$U_B(k)=\sum_{B'} U_{B}^{(B')}(k)$.
%
Energy density is given by
\begin{eqnarray}
\varepsilon&=&
\varepsilon_{mass}+
\varepsilon_{kin}+\varepsilon_{pot} 
\nonumber
\\
&=& 2\sum_{B} \int_0^{k_F^B} \frac{d^3k}{(2\pi)^3}
\left\{ 
M_B+
\frac{\hbar^2 k^2}{2M_B}+\frac 12 U_B(k)\right\} 
\end{eqnarray}
A baryon number density is given as $\rho=\sum_B \rho_B$,
$\rho_B$ being density of component $B$.
Chemical potentials $\mu_B$ and pressure $P$ are expressed as
\begin{eqnarray}
&&\mu_B = \frac{\partial \varepsilon}{\partial \rho_B} \ , 
\label{eq:chem} \\
&& P = \rho^2 \frac{\partial (\varepsilon/\rho)}{\partial \rho_B}
 =\sum_B \mu_B \rho_B -\varepsilon \ .
\label{eq:press}
\end{eqnarray}

In neutron-star matter composed of
$n$, $p$, $e^-$, $\mu^-$, $\Lambda$ and $\Sigma^-$,
equilibrium conditions are given as

\noindent
(1) chemical equilibrium conditions,
\begin{eqnarray}
\label{eq:c1}
&& \mu_n = \mu_p+\mu_e \\
&& \mu_\mu = \mu_e \\
&& \mu_\Lambda = \mu_n \\
&& \mu_{\Sigma^-} =\mu_n + \mu_e 
\label{eq:c2}
\end{eqnarray}
\noindent
(2) charge neutrality,
\begin{eqnarray}
\rho_p = \rho_e +\rho_\mu+\rho_{\Sigma^-}
\end{eqnarray}
\noindent
(3) baryon number conservation,
\begin{eqnarray}
\rho = \rho_n +\rho_p +\rho_\Lambda+\rho_{\Sigma^-}
\label{eq:c3}
\end{eqnarray}
\noindent
When the analytical expressions for energy densities are substituted into 
the chemical potentials (\ref{eq:chem}), the chemical equilibrium conditions 
are represented as equations for densities $\rho_a$
($a=$ $n$, $p$, $e^-$, $\mu^-$, $\Lambda$, $\Sigma^-$).
Then, equations can be solved iteratively, and an energy density and 
a chemical potential are determined for each baryon component.

ESC gives potentials in $S=-1$ and $S=-2$ channels, being designed 
consistently with various data of $Y\!N$ scattering and hypernuclei. 
Important is to determine MPP and TBA parts in channels including hyperons:
MPP's are defined universally in all baryon channels. 
For TBA's in $Y\!N$ channels, 
parameters $V_0$ and $\eta$ in each $Y\!N$ channel are determined
so as to reproduce the related hypernuclear data.
Such a task was performed in \cite{YFYR14,YFYR15}:
Here, the parameter sets in $Y\!N$ channels
are taken from these references, in which
the MPP strengths in $Y\!N$ channels are rather weaker than
the MPP ones in $N\!N$ channels determined in this work.
Though this choice for MPPs is considered to bring about some 
over-estimation of softening effect by hyperon mixing, the conclusion 
for radii of neutron stars in this work is not affected. 

For the EoS softening, the $\Sigma^-$ mixing is more important than
$\Lambda$ mixing, because the electron mass reduces the threshold energy
in the equilibrium condition $\mu_{\Sigma^-} =\mu_n + \mu_e$ in spite of
the $\Sigma^-$ mass larger than $\Lambda$. 
In many RMF approaches, no $\Sigma^-$ mixing occurs due to the condition
of $U_{\Sigma^-}=-(20-30)$ MeV. In our approach assuming the universal
MPP repulsions among all baryons, there appears always $\Sigma^-$ mixing
together with $\Lambda$ and $\Xi^-$ mixing \cite{YFYR15}. However, 
if extra repulsions among $\Sigma nn$ are assumed, 
the $\Sigma^-$ mixing disappears. Namely, there appears no $\Sigma^-$ mixing, 
if repulsive effects for $\Sigma^-$'s are substantially stronger than those 
for nucleons. Such a case can be seen also in \cite{Togashi2}.
In the case of $\Xi^-$ mixing, neglected in this work, the large mass 
of $\Xi^-$ makes the softening effect smaller than the $\Sigma^-$ mixing,
and the effect of $\Xi^-$ mixing on the $M\!R$ relation is small \cite{YFYR15}.

\begin{figure}[tbh]
\includegraphics[width=\columnwidth]{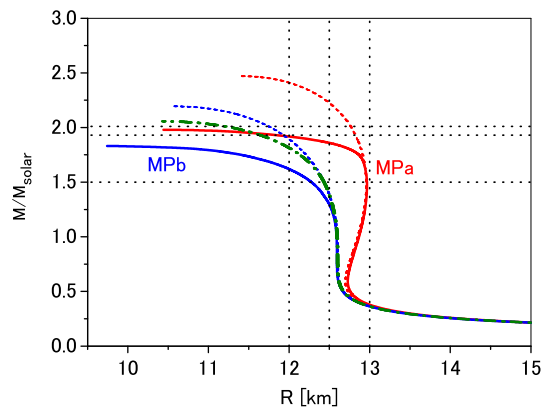}
\caption{(Color online)
Neutron-star masses as a function of the radius $R$.
Solid (dashed) curves are with (without) hyperon 
($\Lambda$ and $\Sigma^-$) mixing for ESC+MPa and ESC+MPb.
Dot-dashed curve for MPb is with $\Lambda$ mixing only.
Also see the caption of Fig.\ref{MRnuc1}.
}
\label{MRhyp}
\end{figure}

Using the EoS of hyperonic nuclear matter, we solve the 
TOV equation to obtain mass-radius relations of neutron stars
in the same way as the previous cases with no hyperon mixing. 
In Fig.~\ref{MRhyp}, 
neutron-star masses are drawn as a function of radius $R$. 
In these figures, solid curves are for MPa and MPb
with hyperon ($\Lambda$ and $\Sigma^-$) mixing, and
dashed curves are obtained without hyperon mixing.
The differences between solid and dashed curves demonstrate
the softening of EoS.
It is found that the maximum mass for MPa is still $2M_\odot$
in spite of remarkable softening of the EoS by hyperon mixing, 
but that for MPb is substantially less than $2M_\odot$.
The dot-dashed curve for MPb is obtained by omitting the 
$\Sigma^-$ mixing, that is including only $\Lambda$ mixing. 
In this case, the maximum mass turns out to become $2M_\odot$ 
owing to the lacking of the large softening effect 
by $\Sigma^-$ mixing. 
Let us add here that the hyperon-onset mass in the solid 
(dot-dashed) curve is $1.65M_\odot$ ($1.51M_\odot$).

Thus, we can consider the two scenarios for the existence
of neutron stars with $2M_\odot$:

\noindent
(1) MPa-type with $\Lambda$ and $\Sigma^-$, where a star mass far larger
than $2M_\odot$ is reduced to $2M_\odot$ by strong softening of EoS
as shown by the solid curve for MPa in Fig.~\ref{MRhyp}.

\noindent
(2) MPb-type with $\Lambda$ mixing only (no $\Sigma^-$ mixing),
where a star mass is kept to be of $2M_\odot$ owing to weak softening 
of EoS as shown by the dot-dashed curve in Fig.~\ref{MRhyp}.

Now, it should be noted that values of $R(1.5M_\odot)$ are 13.0 km
and 12.4 km in the cases of (1) and (2), respectively,
giving the difference of 0.6 km.
As found in the figure, these values are not so affected 
by hyperon mixing, being originated from the MPP strengths
in respective cases. In the MPa case, for instance, the softening
effect of EoS for radii $R$ is found to appear in the region of 
$M>1.8M_\odot$.
Thus, even considering hyperon mixing, we have the same statements 
on the relation between MPP strengths and $R(1.5 M_\odot)$ values 
in previous sections. 
If radii $R$ for $M=(1.4\!\sim\!1.8)M_\odot$ are 
observed with a precision of $\pm 5\%$, 
scenarios (1) and (2) might be discriminated.
Of course, if radii of neutron stars around the maximum mass 
are also observed, it will give more decisive information 
for hyperon mixing.
%

One of the highly prioritized targets observed by NICER is 
the neutron star PSR 0437$-$4715
whose mass is $(1.76\pm0.20)M_\odot$~\cite{PSR0437}.
Our calculated values for $R(1.76M_\odot)$ are as follows:
In ESC+MPa case, we obtain 12.8 km and 12.9 km with and
without hyperon mixing, respectively.
In the ESC+MPb case, we obtain 12.1 km and 12.2 km
with and without hyperon ($\Lambda$) mixing.
Effects of hyperon mixing are still small in the case of
$M=1.76M_\odot$.
When the mass is close to the upper limit of the observed value
$1.96M_\odot$, we have 11.4 km (12.8 km) in ESC+MPa case and 
11.5 km (11.9 km) in ESC+MPb case with (without) hyperon mixing.
On the other hand, in the case of the lower limit of 
$M=1.56M_\odot$ we have 13.0 km (12.4 km) for ESC+MPa (ESC+MPb)
irrelevantly to hyperon mixing.
Thus, our prediction for PSR 0437$-$4715 is summarized as
follows:  The radius is expected to be $11.4\!\sim\!13.0$ km.
If the observed value is $12.8\!\sim\!13.0$ km, the strong softening 
of EoS by hyperon mixing is indicated to be in neutron stars 
of $M>1.8M_\odot$.

\section{Conclusion}
The existence of neutron stars with 2$M_\odot$ gives severe conditions 
for the stiffness of EoS of neutron-star matter.
Though the strong many-body repulsion can make the EoS stiff enough,
the hyperon mixing in neutron-star matter brings about the 
remarkable softening of the EoS. One way to solve this puzzle is to 
consider many-body repulsions working universally among baryons. 
The multi-pomeron potential (MPP) is such a model.
The strength of MPP in nucleon sectors can be determined by fitting 
the observed angular distribution of $^{16}$O+$^{16}$O elastic scattering 
at $E_{in}/A=70$ MeV with use of the G-matrix folding potential.
The neutron-star EoS including MPP contributions is stiff enough to give 
the large neutron-star mass $2M_{\odot}$, which can be obtained with 
no ad hoc parameter for stiffness of EoS in our approach.

The strength of the MPP repulsion plays a role to adjust the stiffness 
of the EoS, changing both mass and radius of a neutron star simultaneously.
Then, values of radii $R$ around a typical mass $1.5M_\odot$ are 
determined by MPP strengths only with almost no effect by hyperon mixing.
On the basis of our analysis using BHF and CVM, we predict
$R(1.5M_\odot)= 12.3\!\sim\!13.0$ km where the width of calculated values
comes mainly from MPP modeling composed of 3- and 4-body repulsions 
or 3-body repulsion only.
We obtain $R(1.5M_\odot)= 12.3\!\sim\!12.4$ km for MPb-type model including 
3-body repulsion only, and $R(1.5M_\odot)= 12.8\!\sim\!13.0$ km
for MPa-type model including 3- and 4-body repulsions.
Precise measurements by NICER for neutron-star radii within $\pm 5\%$
are expected to determine the stiffness of EoS originated from
MPP repulsions.

Information on hyperon mixing can be obtained indirectly from
precise measurements of radii. If $R(1.5M_\odot)$ is larger than
about 12.8 km, remarkable softening of EoS by hyperon mixing has to 
bring about masses of $2M_\odot$. If $R(1.5M_\odot)$ is smaller than
about 12.4 km, the softening of EoS by hyperon mixing has to be weak
in order to keep the maximum mass of $2M_\odot$.

\section*{Acknowledgments}
The author (T.F.) is supported by Grant-in-Aid for JSPS Research Fellow 
(JP15K17661) from the Japan Society for the Promotion of Science.

\end{document}